\documentstyle[aps,prl,array,multicol,psfig]{revtex}
\begin{document}

\title{Multiparticle Landau-Zener problem.}

\author {N.A. Sinitsyn$^1$}
\address{$^1$Department of Physics, Texas A\&M University, College Station,
Texas 77843-4242}

\maketitle
\date{ }
\begin{abstract}
We propose a simple ansatz that allows to generate new exactly
solvable multi-state Landau-Zener models.  It is based on a system
of two decoupled two-level atoms whose levels vary with time and
cross at some moments.
Then we consider multiparticle systems with Heisenberg equations for annihilation operators having
similar structure with Shr\"odinger equation for amplitudes in multistate Landau-Zener models and show that the
corresponding Shr\"odinger equation in multiparticle sector belongs to the multistate Landau-Zener class. This observation allows to 
generate new exactly solvable models from already known ones. 
 We discuss possible applications of the new solutions in the problem of the 
driven charge transport in quantum dots.
\end{abstract}
 \maketitle



\section{Introduction.}

Landau-Zener (LZ) theory \cite{{landau},{zener}} is one of the
most important and influential results in non-stationary quantum
mechanics. Last decade a generalization of LZ theory to more
than two states attracted particular attention due to numerous
applications in atomic and molecular physics \cite{inc},
\cite{nik}, nanomagnets \cite{nm}, Bose-Einstein condensate \cite{Y2} and
systems with avoided band crossings \cite{Gefen}, \cite{Iliescu}.
The multi-state LZ problem (see for example \cite{brand} )
is concerned with finding of the transition amplitudes for a system with
the Hamiltonian, whose matrix form reads:

\begin{equation}
 H=Bt+A,
 \label{mlz}
\end{equation}
where $B$ is a diagonal matrix and the matrices $A$ and $B$ are
independent of time. In its general form this problem is still
unsolved, but a number of exact results for special choices of the
matrices $B$ and $A$ were found \cite{{brand},{deminf1},{bow},{dem33},{zeeman1},{zeeman2},{Y1},{dem3},{sin},{demkov},{usuki}}.

In almost all available exact solutions the transition probabilities
are expressed in  terms of the genuine two-level LZ formula
successively applied at each diabatic level intersection. In other
physical problems such a procedure is often applied as an
approximation. These problems include atomic and molecular
collisions \cite{mol1} and the transitions at crossing of two
Rydberg multiplets of energy levels \cite{inc}.

In this work we find very simple ansatz that generates new
solvable models and may explain the properties of already
known solutions. The main idea employs  the
single-particle Hamiltonian which acts independently in several
two-dimensional subspaces of the Hilbert space. It is worth mentioning that while results
in one-particle sector are trivial, the same
Hamiltonian generates non-trivial
solutions in the many particle spaces. Such a construction is akin to the group-theoretical
method of finding higher irreducible representations as a
symetrized direct product of the fundamental representation. Using 
this method we can study the problem of driven charge transport through a quantum dot and find new solutions in 
multistate LZ theory. Particularly, we  derive the transition probabilities for a four
state LZ problem which is very similar to the four state bow-tie model and for a problem of intersection of two bands of parallel levels.

This article is organized as follows: in section II we show how already known solutions of LZ models can generate new exactly solvable models
with the Hamiltonian (\ref{mlz}). We demonstrate how the exact solution for two independent two level systems can generate a new solution of a
four-state LZ model. In III we generalize Demkov-Osherov solution to the case of many particles and use the result for derivation of
 master equations that describe a driven charge transport in quantum dots. In IV we provide an example of a solvable model that can be 
 generated from the Demkov-Osherov solution. 

\section{Bosonic multi-state LZ models.}

Lets consider a Hamiltonian that describes the interaction of four bosonic
fields $\hat{a},\, \hat{b},\, \hat{c},\, \hat{d}$:

\begin{equation}
      \begin{array}{l}
      \hat{H}=(\beta_1 t + E_1) \hat{a}^+ \hat{a} +(\beta_3 t +E_3) \hat{d}^+
      \hat{d} + (\beta_2 t+E_2) \hat{c}^+ \hat{c} + (\beta _4 t+E_4)
       \hat{b}^+ \hat{b}+ \\
      g(\hat{a}^+ \hat{b} +\hat{b}^+ \hat{a}) +\gamma (\hat{c}^+ \hat{d} + \hat{d}^+ \hat{c})
      \end{array}
      \label{ham1}
      \end{equation}
This Hamiltonian depends explicitly on time and conserves the
total number of particles in the system. Therefore it can be
considered independently in subspaces with fixed total number of
particles. Let $|0>$ be the vacuum state. The Hamiltonian
(\ref{ham1}) describes the evolution of two disjointed systems. However, being projected onto the
2-particle sector, its matrix form looks less trivial. The complete
two-particle sector is the 10-dimensional Hilbert space spanned
onto direct products of any two single-particle states. The
four-dimensional subspace $R_4$ of the 2-particle sector spanned
onto vectors:
\begin{equation}
      \begin{array}{l}
      |1>=\hat{a}^+ \hat{c}^+ |0> \\
      |2>=\hat{a}^+ \hat{d}^+ |0> \\
      |3>=\hat{d}^+ \hat{b}^+ |0> \\
      |4>=\hat{c}^+ \hat{b}^+ |0>
      \end{array}
      \label{states}
\end{equation}
is invariant with respect to the action of the Hamiltonian
(\ref{ham1}). Hence, if the initial state belongs to this
subspace, the state vector at any time remains in $R_4$:

\begin{equation}
      |\psi(t)\rangle = c_1 (t) |1\rangle +c_2 (t) |2\rangle
      + c_3 (t) |3\rangle +c_4 (t) |4\rangle
      \label{ev}
\end{equation}
In the basis (\ref{states}) the Hamiltonian (\ref{ham1}) has the following 4x4 matrix form:

\begin{equation}
H=\left(
\begin{array}{llll}
(\beta _1+ \beta_2)t +(E_1+E_2) & \,\, \gamma & 0 & g \\
\gamma & (\beta_1+\beta_3)t+(E_1+E_3) & g & 0\\
0 & g & (\beta_3+\beta_4) t +(E_3+E_4) & \gamma \\
g & 0 & \gamma & (\beta_2+\beta_4) t +(E_4+E_2)
\end{array}
\right)
\label{ham2}
\end{equation}

The problem described by the Hamiltonian (\ref{ham2}) belongs to the multistate Landau-Zener class (\ref{mlz}).
 
We should point out that it cannot be mapped on the already known exactly solvable multistate LZ models.



\begin{figure}
\psfig{figure=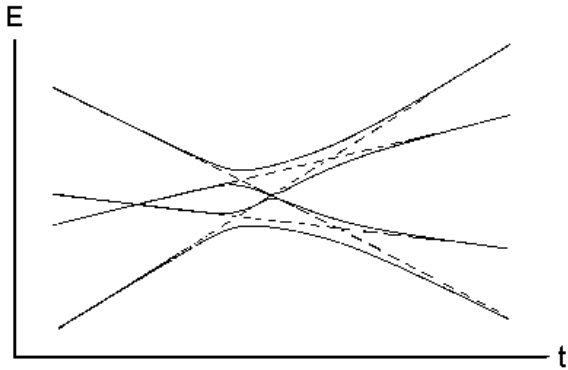,width=12.6cm}
\vskip-2in
\caption{Time dependence of the adiabatic energies (solid lines) and diagonal elements (dashed lines) of the Hamiltonian (\ref{ham2}). The
choice of parameters is $\beta_1=5, \, \beta _2=-3, \, 
\beta _3 = 0, \, \beta_4=-1.5, \, E_1=3, \, E_2=0.5, \, E_3=-2, \, E_4=-1.5, \, g=1, \, \gamma=1.5$ }
\label{Fig1}
\end{figure}


In $Fig.1$ we show the adiabatic energies of the Hamiltonian (\ref{ham2}) as functions of time for typical choice of parameters.
 It is clearly seen that there is a point of an
exact adiabatic level crossing (diabolic point) on the figure. 

In the Heisenberg representation the evolution equations decouple
into two pairs of equations for bosonic operators:

\begin{equation}
\begin{array}{l}
i \dot{\hat{a}}=(\beta_1 t+E_1) \hat{a} + g \hat{b} \\
i \dot{\hat{b}}=(\beta_4 t+ E_4) \hat{b} + g \hat{a}
\end{array}
\label{eqq1}
\end{equation}
and
\begin{equation}
\begin{array}{l}
i \dot{\hat{c}}=(\beta_2 t +E_2) \hat{c} + \gamma \hat{d} \\
i \dot{\hat{d}}= (\beta_3 t +E_3) \hat{d} + \gamma \hat{c}
\end{array}
\label{eqq2}
\end{equation}
Let $\hat{a}_0,\hat{b}_0,\hat{c}_0,\hat{d}_0$ denote the operators
$\hat{a}, \hat{b}, \hat{c}, \hat{d}$ at the initial moment of
evolution. Then the solutions of equations (\ref{eqq1}) and
(\ref{eqq2}) are:

\begin{equation}
\begin{array}{l}
\hat{a}(t) = S_{11} (t) \hat{a}_0 + S_{12} (t) \hat{b}_0 \\
\hat{b} (t) = S_{21}(t) \hat{a}_0 + S_{22} (t) \hat{b}_0
\end{array}
\label{ab1}
\end{equation}
\begin{equation}
\begin{array}{l}
\hat{c}(t) = S_{11}^{\prime} (t) \hat{c}_0 + S_{12}^{\prime} (t) \hat{d}_0 \\
\hat{d} (t) = S_{21}^{\prime}(t) \hat{c}_0 + S_{22}^{\prime} (t)
\hat{d}_0
\end{array}
\label{ab2}
\end{equation}
Here $S_{ij}$ and $S_{ij}^{\prime}$ are the matrix elements of the
evolution operators for (\ref{eqq1}) and (\ref{eqq2}),
respectively. Due to the linearity they are the same for the operator
and numerical functions obeying these differential equations.
Hence, we can extract them directly from the solution of the
two-state LZ problem. For the evolution from $t=-\infty$ to
$t=+\infty$ their squares of modulus are:
 \begin{equation}
 \begin{array}{l}
 p_1\equiv |S_{11}|^2=|S_{22}|^2=e^{-2\pi g^2/|\beta_1-\beta_4|} \\
 q_1\equiv |S_{12}|^2 = |S_{21}|^2 = 1-p_1 \\
 p_2\equiv |S_{11}^{\prime}|^2=|S_{22}^{\prime}|^2=e^{-2\pi \gamma ^2/|\beta_2 - \beta_3|} \\
 q_2\equiv |S_{12}^{\prime}|^2 = |S_{21}^{\prime}|^2 = 1-p_2 \\
 \end{array}
 \label{pq}
 \end{equation}
Returning to the four-state LZ problem in the two-particle sector
considered earlier, we first note that each state $|\gamma\rangle$
of this subspace is the direct product of states from two
independent subspaces of the one-particle sector
$|j\rangle=|\alpha_{j}\rangle\bigotimes|\mu_{j}\rangle$,
$\alpha_{j}=1,2; \mu_{j}=3,4$ (note that here 1,2,3,4 enumerate
single-particle state, for example $|1\rangle=a^+|0\rangle$). The
evolution matrix is also the direct product of evolution matrices
in the independent subspaces of the one-particle sectors:
$U(t)=U_{\alpha}(t)\bigotimes U_{\mu}(t)$. Therefore transition
matrix elements and probabilities $P_{ij}$ in the considered
subspace are factorized:
\begin{equation}
P_{ij}=p_{\alpha_i\alpha_j}p_{\mu_i\mu_j} \label{factorization}
\end{equation}
In terms of the LZ probabilities for two-level problems introduced
earlier the transition probability matrix $P$, whose elements are
defined by equation (\ref{factorization}), reads:


 \begin{equation}
P= \left(
 \begin{array}{llll}
 p_1 p_2 & p_1 q_2 & q_1 q_2 & p_1 q_2 \\
 p_1 q_2 & p_1 p_2 & q_1 p_2 & q_1 q_2 \\
 q_1 q_2 & p_2 q_1 & p_1 p_2 & p_1 q_2 \\
 q_1 p_2 & q_1 q_2 & p_1 q_2 & p_1 p_2
 \end{array}
 \right)
 \label{fin1}
 \end{equation}
This result does not depend on the parameters $E_i$. It is interesting
that scattering matrices $S_{ij} (t)$ and $S_{ij}^\prime (t)$ are
known for any $t$ \cite{zener} which make it possible to find the
evolution operator at any time in the Schr\"odinger
representation.

\section{Driven charge transport through quantum dots.}

  In similar fashion to the previous section the fermionic systems can
 lead to Heisenberg equations for annihilation operators that have the same
 structure as Shr\"odinger equation for amplitudes for some exactly solvable
 multistate Landau-Zener model. As we will show, in the Shr\"odinger representation such a Fermi system with fixed number of particle is equivalent
 to a new solvable multistate Landau-Zener model. 
The models that we will examine correspond to the driven charge transport in nanostructures.  

Consider a quantum dot coupled to an external reservoir like  the system shown in $Fig 2$. Lets consider that initially some of the reservoir  
 energy levels are filled with electrons, the others are
empty. Lets assume the dot has only one electron bound state whose
energy in real semiconductors can be regulated by the gate
voltage; therefore the variation of the gate voltage with time generates time
dependence of the dot's electronic level.

\begin{figure}
\psfig{figure=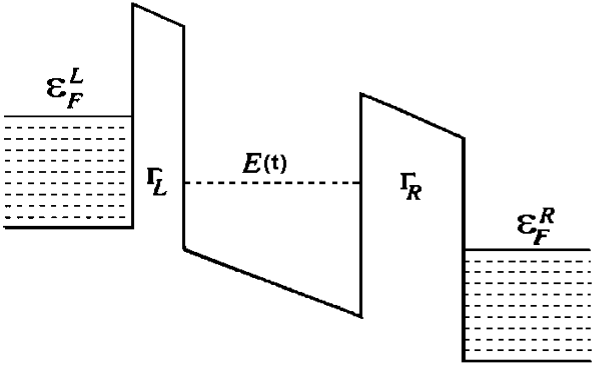,width=11.6cm,clip=}
\caption{A single energy level in a potential well coupled to two leads at zero temperature. 
Electron states in leads are filled up to Fermi energies,
that can be different in right and left leads.}
\label{Fig2}
\end{figure}
The Hamiltonian of the electrons in the dot and reservoir reads:
\begin{equation}
H = \sum \limits_{n=1} ^{N} E _n \hat{c} _n^{+} \hat{c} _n +E(t)
\hat{c}_0^{+} \hat{c}_0 +\sum \limits_{n} g _n( \hat{c} _n^{+} \hat{c}_0 + \hat{c}_0
 ^{+} \hat{c} _n)
\label{h1}
\end{equation}
Here $\hat{c_0}$ is the fermionic operator that annihilates the
electron on the dot level and $\hat{c}_n$ is the annihilation
operators for the level $E_n$ of the reservoir; $E(t)$ is the
time-dependent energy of the dot state. In our treatment the last term in (\ref{h1}) 
describes the tunneling between the leads and the single level in the quantum dot.  We ignore all interactions among electrons except the one
due to Pauli principle.

Similar time-dependent single-particle problems for quantum dots
have been already considered in \cite{dot2}. Though our system is simplified but rather it is interesting because it
has an exact solution.

In the context of LZ theory, we approximate the dot energy by a
linear function of time: $E(t)=\beta t$. The Heisenberg operator
equations corresponding to Hamiltonian (\ref{h1}) are:

\begin{equation}
\begin{array}{l}
 i\dot{\hat{ c }}_0 = \beta t \hat{ c }_0 + \sum \limits _{n}^{}  g_n \hat{ c }_n \\
i \dot {\hat{ c }}_n = E_n \hat{ c }_n + g _n \hat { c }_0
\end{array}
\label{se1}
\end{equation}
Due to the linear structure of these equations the
solution can  be formally written in the matrix form:
\begin{equation}
\vec{ \hat{c}}(t) = \hat { S } (t) \vec { \hat{c} }(t_0)
\label{sol1}
\end{equation}
where $\vec{\hat{c}}=(\hat{c}_0,\hat{c}_1,\ldots ,\hat{c}_n)$

As in the previous section, the evolution matrix $\hat{S}(t)$ is
completely determined by the coefficients of the differential equations
(\ref{se1}) and is the same for operator and c-function solutions.
Hence, it is enough to solve (\ref{se1}) with all operators
replaced by c-functions. Such a system of equations coincides with
that of the Demkov-Osherov model \cite{demkov}. The latter
provides transition amplitudes for a single energy level crossing
an energy band consisting of time-independent levels. 

 In Demkov-Osherov model the Shr\"odinger equation for the
amplitudes of different quantum states can be written as follows:
  \begin{equation}
      i\left( {\begin{array}{*{20}c}
   {\dot a_0(t)}  \\
   {\dot a_1 (t)}  \\
    \vdots   \\
   {\dot a_n (t)}  \\
\end{array}} \right) = \left( {\begin{array}{*{20}c}
   {\beta t} & {\gamma _1 } &  \cdots  & {\gamma _n }  \\
   {\gamma _1^* } & {\alpha _1 } &  \cdots  & 0  \\
    \vdots  &  \vdots  &  \ddots  &  \vdots   \\
   {\gamma _n^* } & 0 &  \cdots  & {\alpha _n }  \\
\end{array}} \right ) \left ( {\begin{array}{*{20}c}

   {a_0(t)}  \\
   {a_1 (t)}  \\
    \vdots   \\
   {a_n (t)}  \\
\end{array}}\right )
  \label{d1}
  \end{equation}
where $\alpha_k$ are ordered as follows
$\alpha_1<\alpha_2<...<\alpha_{N_a}$ (assuming that none  two of the
$\alpha_k$ are equal; we also assume for definiteness that
$\beta>0$).

The absolute values of the $S$-matrix components 
($S_{kl}=|a_k^{(l)}(\infty)|/a_l^{(l)}(-\infty)|)$ \cite{demkov} are:

\begin{equation}
\begin{array}{l}
 S_{00}  = e^{ - \pi (z_1  +  \ldots  + z_n )} \\
 S_{0l}  = (1 - e^{ - 2\pi z_l } )^{1/2} e^{ - \pi (z_{l + 1}  +  \ldots  + z_n )} \\
 S_{k0}  = e^{-\pi (z_1  +  \ldots  + z_{k - 1} )} (1 - e^{ - 2\pi z_k } )^{1/2} ,(k = 1, \ldots ,n) \\
 S_{kl}  = 0,(1 \le k < l)  \\
 S_{ll}  = e^{ - \pi z_l }  \\
 S_{kl}  = (1 - e^{ - 2\pi z_l } )^{1/2} e^{ - \pi (z_{l + 1}  +  \ldots  + z_{k - 1} )} (1 - e^{ - 2\pi z_k } )^{1/2}, (k>l)
\end{array}
\label{d10}
\end{equation}

(where the index $l=1\ldots n$  and $z_k=|\gamma_k|^2/\beta$)

 The probabilities to
find an electron on a particular $n$-th level are. 
\begin{equation}
\begin{array}{l}
P_n=<\hat{c}_n^{+} (t \rightarrow +\infty) \hat{c}_n (t\rightarrow
+\infty)>= \sum _{n_1} \sum_{n_2} S_{n n_1}^*  S_{n n_2}
<\hat{c}_{n_1}^{+} (t \rightarrow -\infty) \hat{c}_{n_2}
(t \rightarrow -\infty)>= \\
=\sum \limits _{n_f}{} |S_{n,n_f}|^2
\end{array}
\label{pnum1}
\end{equation}
where $S_{ij} = S_{ij} (t \rightarrow +\infty)$
and the summation is taken over the initially filled states only.
The scattering matrix elements $S_{n,n_f}$ are given in (\ref{d10}).
 If the band of electron states in the external system is
continuous then it is reasonable to use the approximation,
in which $g(E)=g_n \rightarrow 0$ while the value $\Gamma(E) = 2
\pi \rho (E) |g(E)|^2 $ is kept finite. Here $\rho (E)$ is the
density of states in the band and 
 the elements of scattering matrix become
$|S_{0l}|^2=\frac{2 \pi g_l^2}{\beta} \exp{\int \limits _{E_l}^{E_n} \frac{-\Gamma(E)}{\beta}dE}$

Now lets consider a dot that is connected to two leads. The left
lead is characterized by the coupling function $g_L (E)$ and the densities
$\rho _L ^f (E)$, $\rho _L ^e (E)$ where f and e refer to the filled
and empty states in the left lead  ($\rho _L (E)=\rho _L ^f (E)+\rho _L ^e (E)$), analogously we can define the
 quantities $g_R (E)$,
$\rho _R ^f (E)$, $\rho _R ^e (E)$ for the right lead.
Moreover it is more convenient to introduce the following notations: $\Gamma _L^f
(E)=2 \pi \rho_L^f (E) |g_L(E)|^2$, $\Gamma _L^e (E)=2 \pi
\rho_L^e (E) |g_L(E)|^2$, $\Gamma _R^f (E)=2 \pi \rho_R^f (E)
|g_R(E)|^2$, $\Gamma _R^e (E)=2 \pi \rho_R^e (E) |g_R(E)|^2$,
$\Gamma^f = \Gamma _R ^f +\Gamma _L ^f$, and $\Gamma ^e = \Gamma
_R ^e +\Gamma _L ^e$.

 If the dot state was initially empty and if
this state crosses the region from energy $E_1$ to energy $E_2$,
then in the continuous approximation (\ref{pnum1}) leads to
the following probability for the dot level to be finally filled after all Landau-Zener transitions:

\begin{equation}
p_f(E_2)=P_0=\int \limits _{E_1} ^{E_2} \frac{\Gamma ^f
(E^{\prime})} {\beta} \exp{\left[-\frac{1}{\beta} \int \limits
_{E^{\prime}} ^{E_2} ( \Gamma ^f (E) + \Gamma ^e
(E))dE\right]}dE^{\prime} \label{p0}
\end{equation}
If the dot level was initially filled, it is necessary to add
$|S_{00}|^2= e^{-\frac{1}{\beta} \int\limits _{E_1} ^{E_2}
( \Gamma ^f (E) + \Gamma ^e (E))dE}$ to (\ref{p0}). One can check that the result
(\ref{p0}) is the solution of the following system of differential
equations:

\begin{equation}
\begin{array}{l}
 \beta \frac{d p_f (E)}{dE} = - \Gamma^e (E) p_f (E)+ \Gamma ^f (E) p_e (E) \\
 \beta \frac{d p_e (E)}{dE} = - \Gamma^f (E) p_e (E)+ \Gamma ^e (E) p_f (E)
\end{array}
\label{dp}
\end{equation}
here $p_e (E)=1-p_f (E)$ is the probability that the dot level
will be empty when it has energy $E$. The equation for the charge
that is transferred to the right lead  can be derived in a similar way

\begin{equation}
 \frac{d Q (E)}{dE} =(e/ \beta )(  \Gamma _R^e (E) p_f (E) - \Gamma _R ^f (E) p_e (E))
\label{dq}
\end{equation}
  Note that equations (\ref{dp}),(\ref{dq}) were derived from the
exact solution of the problem with microscopic Hamiltonian (\ref{h1}) rather
than from random phase approximation or other type of
phenomenology.

Let us calculate the total charge transferred through the dot from
the left lead to the right lead at zero temperature and a fixed bias
that leads to a difference of Fermi energies in the left and in
the right leads. Lets assume that the dot level was initially much lower than
both Fermi levels and it was filled. Then the energy of this state
grows linearly with time crossing both
Fermi levels  during the evolution. Since transitions will proceed
presumably when the dot level is between Fermi energies of the leads, we can apply the following
approximations: $\Gamma _L ^f (E) = \Gamma _L (1 -\theta (
E-\epsilon _F^L))$, $\Gamma _L ^e (E) = \Gamma _L \theta (
E-\epsilon _F^L)$, $\Gamma _R ^f (E) = \Gamma _R (1 -\theta (
E-\epsilon _F^R))$ and $\Gamma _R ^e (E) = \Gamma _R \theta (
E-\epsilon _F^L)$ with $\Gamma _R$ and $\Gamma _L$ are constant. To find the total charge that is
transferred to the right lead we formally put the final dot state
energy equal to infinity in the solution of the 
equations (\ref{dp}) and (\ref{dq}). In the result the total charge transfered to the right lead is

 \begin{equation}
 Q=e\left[\frac{\Gamma _R \Gamma _L}{(\Gamma _R + \Gamma _L)}
 (\frac{\epsilon _F^L -\epsilon _F^R}{\beta} ) + \frac{\Gamma _R }{(\Gamma _R +\Gamma _L)}\right]
 \label{Q}
 \end{equation}
Clearly at  $\epsilon _F^L =\epsilon _F^R$ we find $Q=e\Gamma _R /(\Gamma _R +
\Gamma _L)$, which can be interpreted as the electron charge $e$ multiplied by the
probability for the electron that is initially placed into the dot to transfer
to the right lead.

\section{Solvable model of bands crossing.}

In this section we will construct a new solvable LZ model employing the
fermionic Hamiltonian. The Hamiltonian (\ref{h1}) projected onto
the $k$-particle sector generates the evolution in the Hilbert
space of dimensionality $(N+1)!/(k!(N+1-k)!)$. If we assume that
the single-particle Hamiltonian laying in the background is the
same as that of the Demkov-Osherov model, then all such models are
reducible to this single-particle one.

Generalized Landau-Zener models that deal with intersections of
bands of parallel levels are important in many applications such as in driven tunneling of
nanomagnets coupled to nuclear spins \cite{nm} and in driven charge transport in quantum dots \cite{usuki}.

Up to now only two exact solutions of this type were known: 
Demkov-Osherov solution and the case of the infinite 
number of states in bands that equally interact with states of another band \cite{deminf}, \cite{deminf1}. For an important
case of a finite number of states in bands that is not equal to
unity exact solutions for all transition probabilities have not been found yet though the absence of counterintuitive transitions was
analytically proved \cite{usuki}. Nearly-exact solution valid in the quasidegeneracy approximation was found and investigated in \cite{Y1}.
We will show that our method can be used to generate exactly solvable models with interband transitions.

Lets consider a system of two Fermi particles with the Hamiltonian
\begin{equation}
H =  E _1 \hat{b}^+ \hat{b} +E_2 \hat{c}^+ \hat{c}  +
t \hat{d}^{+} \hat{d} +  g _1(\hat{a}^{+} \hat{d} + \hat{d} ^{+} \hat{a})
+  g _2(\hat{b}^{+} \hat{d} + \hat{d} ^{+} \hat{b})+
g _3(\hat{c}^{+} \hat{d} + \hat{d} ^{+} \hat{c})
\label{h2}
\end{equation}
Let $E_2>E_1>0$.  As we demonstrated previously, the solution of the
operator evolution equation can be written in the form:

\begin{equation}
\left(
\begin{array}{l}
\hat{d} (t)\\
\hat{a} (t)\\
\hat{b} (t)\\
\hat{c}(t)
\end{array} \right)
= S(t,t_0) \left(
\begin{array}{l}
\hat{d} (t_0)\\
\hat{a} (t_0)\\
\hat{b} (t_0)\\
\hat{c}(t_0)
\end{array} \right),
\label{vec2}
\end{equation}
where $S(t,t_0)$ is the matrix  of evolution for a 4-state
Demkov-Osherov model. Lets restrict the Hilbert space to the
subspace of only two particles. It includes six states:
$|1>=\hat{d}^+\hat{a}^+|0>$, $|2>=\hat{d}^+\hat{b}^+|0>$,
$|3>=\hat{d}^+\hat{c}^+|0>$,  $|4>=\hat{a}^+\hat{b}^+|0>$,
$|5>=\hat{a}^+\hat{c}^+|0>$ and $|6>=\hat{b}^+\hat{c}^+|0>$.
Similarly to the bosonic case, this subspace is invariant during the
evolution process. The Hamiltonian restricted to this subspace has the
following matrix form:
\begin{equation}
H=\left(
\begin{array}{llllll}
t   & 0     & 0     & -g_2 & -g_3  & 0         \\
0   & t+E_1 & 0     & g_1  & 0     & -g_3      \\
0   & 0     & t+E_2 & 0    & g_1   & g_2       \\
-g_2& g_1   & 0     & E_1  & 0     & 0         \\
-g_3& 0     & g_1   & 0    & E_2   & 0         \\
0   & -g_3  & g_2   & 0    &   0   & E_1+E_2
\end{array}
\right)
\label{h3}
\end{equation}

\begin{figure}
\psfig{figure=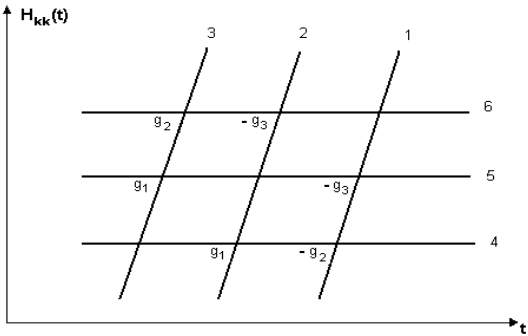,width=8.6cm,clip=}
\caption{ Diagonal elements of the Hamiltonian (\ref{h3}) as functions of time. 
  } \label{Fig3}
\end{figure}

Let $P_{ij}$ ($i,j=1,...,6$) be the probability to transit from
the state $j$ to the state $i$ after the band crossing.
The transition probabilities can be expressed in terms of
the fermi-operators in the Heisenberg representation at
$t\rightarrow\infty$.

\begin{equation}
\begin{array}{l}
P_{1n}=<n|\hat{a}^+\hat{a} \hat{d}^+ \hat{d} |n> \\
P_{2n}=<n|\hat{b}^+\hat{b} \hat{d}^+ \hat{d} |n> \\
P_{3n}=<n|\hat{c}^+\hat{c} \hat{d}^+ \hat{d} |n> \\
P_{4n}=<n|\hat{a}^+\hat{a} \hat{b}^+ \hat{b} |n> \\
P_{5n}=<n|\hat{a}^+\hat{a} \hat{c}^+ \hat{c} |n> \\
P_{6n}=<n|\hat{b}^+\hat{b} \hat{c}^+ \hat{c} |n> \\
\end{array}
\label{pmn}
\end{equation}
Substituting (\ref{vec2}) into (\ref{pmn}) and employing the elements of
the evolution matrix from (\ref{d10}) we get the following
result:
\begin{equation}
P=\left(
\begin{array}{llllll}
p_2 p_3 &q_1 q_2 p_3 &q_1 q_3     &p_1 q_2 p_3 &p_1 q_3     &0       \\
0       &p_1 p_3     &p_1 q_2 q_3 &q_1 p_3     &q_1 q_2 q_3 &p_2 q_3 \\
0       &0           &p_1 p_2     &0           &q_1 p_2     &q_2     \\
q_2     &q_1 p_2     &0           &p_1 p_2     &0           &0       \\
p_2 q_3 &q_1 q_2 q_3 &q_1 p_3     &p_1 q_2 q_3 &p_1 p_3     &0       \\
0       &p_1 q_3     &q_2 p_3 p_1 &q_1 q_3     &q_1 q_2 p_3 &p_2 p_3
\end{array}
\right)
\label{fin2}
\end{equation}
where

\begin{equation}
\begin{array}{l}
p_i=e^{-2 \pi |g_i|^2 } \\
q_i = 1-p_i, \,\,\, (i=1,2,3)
\end{array}
\label{pq}
\end{equation}

\section{Conclusions.}

In conclusion, we presented the procedure that generates new
exactly solvable multi-state LZ models. Some of them are useful
for the description of driven charge transport in quantum dots and
driven tunneling in nanomagnets. As an example, we derived two
new solvable models and found the transition probability matrices
for them.

There have been three known classes of solvable multi-state LZ models
that provide transition probabilities for a finite number of
states:
\begin{enumerate}
\item{The Demkov-Osherov model.}
\item{The $SU(2)$ symmetry class that deals with an arbitrary spin
in external magnetic field with the following Hamiltonian:

\begin{equation}
  \hat{H}=t\hat{S}_z + g \hat{S}_x
  \label{spin}
  \end{equation}
}

\item{The generalized bow-tie model that treats the case when two levels are parallel while the other levels intersect at one point
between the parallel ones.}
\end{enumerate}
This list can be extended with different generalizations of these
models to the case of degenerate states. For example, it is
possible to solve the LZ model for two degenerate levels by
changing basis in such a way that all equations decouple into
independent two state Landau-Zener transitions. It is worth mentioning that sometimes a few elements of transition probability
matrix can be found while the others remain unknown \cite{brand}, \cite{usuki}.

 All these models
provide very simple results. For example, transition probabilities
in the Demkov-Osherov model coincide with those taken from successive application of the two state Landau-Zener formula. The same is true for the
generalized bow-tie model. Finally in all models the transition
probabilities are simple polynomials of $z_k = \exp(-\pi
|g_k|^2)$. This fact gives a strong feeling that there should be a
common symmetry in the background of all these models. Our results demonstrate the same
properties and we know that the reason for this was the symmetry that makes the Hamiltonian equivalent
in some sense to the one for a much simpler problem.

 We did not study deeply the relations between the known models and our new solutions but there are indications that
such a relation can exist. For example
the
model (\ref{ham2}) and the generalized bow-tie model are very
similar. At $g=\gamma$ and $\beta _1=-\beta_2, \, \beta _3 = -\beta _4$ the Hamiltonian (\ref{ham2}) belongs to the
class of generalized bow-tie models. 
Also we note that models of $SU(2)$ class \cite{zeeman1} can be
derived from the following bosonic Hamiltonian:

\begin{equation}
\hat{H} = t \hat{a}^+ \hat{a} - t \hat{b}^+ \hat{b} + g (\hat{a}^+ \hat{b} + \hat{b}^+ \hat{a})
\label{bh}
\end{equation}

In the single-particle sector the Hamiltonian (\ref{bh}) leads to
the simple two-state LZ model. In the $N$-boson sector the
Schr\"odinger equation for diabatic states coincides with ones for
a spin $S=N/2$ in magnetic fields. This construction is an
application of the Schwinger bosons \cite{schwinger} to the LZ
problem.

 It is interesting to
check what models can be reduced to decoupled two level systems.
Probably this can be done using of the group representation
theory.

\begin{acknowledgments} This work was supported by NSF under the
grants DMR 0072115 and DMR 0103068 and by DOE under the grant
DE-FG03-96ER45598. I thank M.A.Kayali and N.Prokof'ev for important remarks
and I am very grateful to V.L. Pokrovsky for
the encouragement, discussion and critical reading of this text.
\end{acknowledgments}

\end{document}